\documentclass[12pt]{article}

\usepackage{bm}
\usepackage{amssymb}
\usepackage{enumerate}

\usepackage{graphicx} 
\begin{document}


\title{\bf Implications of  last NA64 results  and the electron $g_{e}-2$ 
anomaly for the  X(16.7) boson survival}

\author{N.V.~Krasnikov$^{1,2}$\footnote{E-mail: Nikolai.Krasnikov@cern.ch} 
\\
$^{1}$ Institute for Nuclear Research of the Russian Academy of Sciences, Moscow
\\
$^{2}$ Joint Institute for Nuclear Research, Dubna}




\date{        } 
\maketitle

\begin{abstract}
We point out that last NA64 bound on coupling constant of hypothetical $X(16.7 MeV)$ 
vector boson with  electrons  plus  the recent   value of the anomalous electron magnetic moment 
exclude at 90 \% C.L. purely vector or axial vector couplings of  $X(16.7)$  boson with electrons. 
Models with nonzero $V \pm  A$ coupling constant with electron
survive  and  they can   explain  
both the electron  and muon $g - 2$ anomalies.   

\end{abstract}

\newpage



The ATOMKI experiment \cite{ATOMKI} has  found   $6.8~\sigma$ excess of events in 
the invariant mass distribution of $e^+e^-$ pairs produced in the nuclear transition of excited 
$^8Be^*$ to its ground state. Recently these results have been confirmed \cite{ATOMKI1}. 
The ATOMKI result can be interpreted as the emission of a new protophobic 
gauge boson $X$ with a mass of $16.7~MeV$ followed by its $X \rightarrow e^+e^-$ decay 
assuming that the X boson has  vector coupling to electrons in the range
$3 \times 10^{-4} \leq \epsilon_e \leq 1.4 \times10^{-3}$ and the 
lifetime $10^{-14} \leq \tau_X \leq 10^{-12}~s$ \cite{Feng}. 
In the model 
proposed in \cite{Feng} X-boson interacts with $B$ or $B-L$ current and the electromagnetic current, 
so  its interaction with electrons is vector like. 
In ref.\cite{NA64} NA64 collaboration has excluded the values 
$1.3 \times 10^{-4} \leq \epsilon_e \leq 4.2\times10^{-3}$. Recently 
NA64 collaboration \cite{NA64last} has published new results and excluded the values 
\begin{equation}
 \epsilon_e \leq 6.8 \times10^{-4} \,
\end{equation}
for the $X(16.7)$ boson model with vector interaction. 
 Only the values 
$6.8 \times 10^{-4} \leq \epsilon_e \leq 1.4 \times10^{-3}$ are not excluded.
The most general renormalizable interaction of vector $X(16.7)$ boson with electrons 
has the form
\begin{equation}
L_{Xe} = -eX^{\mu}[\epsilon_{Ve}\bar{e}\gamma_{\mu}e + \epsilon_{Ae}\bar{e}\gamma_{\mu}\gamma_5e] \,.
\end{equation}
The  $X(16.7)$ boson contribution  to the anomalous electron moment for $m_X  \gg m_e$  
is \cite{MOMENT}
\begin{equation} 
\frac{(g_e - 2)}{2}  = \frac{\alpha}{3\pi}(\frac{m_e}{m_{X}})^2[\epsilon^2_{Ve} - 5\epsilon^2_{Ae}] \,.
\end{equation}
Recent precise determination of the fine structure constant, $\alpha$ allowed to 
determine \cite{alpha} 
\begin{equation}
\alpha^{-1}(Cs) = 137.035999046(27) \,.
\end{equation}
The use of the value (4) for the fine structure constant $\alpha$ leads to the prediction 
for the anomalous electron  magnetic moment \cite{13, 15}
with the $2.4~\sigma$ negative discrepancy 
with   the measured electron magnetic moment, namely \cite{Marciano} 
\begin{equation}
\Delta a_e \equiv a^{exp}_e - a^{SM}_e = (-0.87 \pm 0.36)\times 10^{-12} \,,
\end{equation}
where $a_e \equiv (g_e-2)/2$. 

In this  note we point out that last NA64 bounds on coupling constant of hypothetical $X(16.7)$ 
vector boson with  electrons  plus  the recent   value  
(5) of the anomalous electron magnetic moment 
exclude at 90 \% C.L. purely vector and axial vector couplings of  $X(16.7)$  boson with electrons. 
Models with nonzero $V \pm  A$ coupling constant with electron
survive and  moreover they can explain  
both the electron  and muon $g - 2$ anomalies.

Consider at first the case of  vector interaction (2), i.e.  $\epsilon_{Ae} = 0$. 
There are several methods  to obtain upper bound  on the $\Delta a_e $ from  the value (5) of 
the electron $\Delta a_e$ in the assumption that new physics contribution to $\Delta a_e$ is non negative    that is valid for the model with vector interaction of the X(16.7) 
with electron. In Cousins-Feldman method \cite{Cousins}\footnote{Here we  assume that $\Delta a_e$ is random variable which is distributed as 
$\Delta a_e =  \frac{1}{\sqrt{2\pi}\sigma} exp[-\frac{(\Delta a_e -(\Delta a_e)_0)^2}{2\sigma^2}]  $ with $(\Delta a_e)_0) \geq 0$ and 
$\sigma = 0.36\cdot 10^{-12}$.} 

as it 
follows from Table 10 of ref.\cite{Cousins} at 90 \% C.L.
\begin{equation}
\Delta a_e \leq \Delta a_e(C.F.) = 0.12 \cdot 10^{-12}\,,
\end{equation} 
\begin{equation}
\epsilon_{eV} \leq 4.1 \cdot10^{-4} \,.
\end{equation}
In Bayes method the upper bound on $\Delta a_e$ at  90 \% C.L. is determined from the equation
\begin{equation}
\frac{\int_0^{\Delta a_e(B)}F(\mu|\mu_0, \sigma)d\mu}{\int_0^{\infty}F(\mu|\mu_0, \sigma)d\mu} = 0.9 \,,
\end{equation}
\begin{equation}
F(\mu|\mu_0, \sigma) = \frac{1}{\sqrt{2\pi}\sigma} exp[-\frac{(\mu-\mu_0)^2}{2\sigma^2}] \,,
\end{equation}
where $\mu_0 = -0.87$ and $\sigma = 0.36$. 
Numerically we find that at 90 \% C.L.  
\begin{equation}
\Delta a_e \leq \Delta a_e(B)  = 0.27 \times 10^{-12}  \,,
\end{equation}
\begin{equation}
\epsilon_{eV}  \leq 6.2 \cdot10^{-4} \,.
\end{equation}

The Bayes value (11)  for  $\epsilon_{eV}$   upper  limit    
is by factor 1.5 higher the Feldman-Cousins  value (7). 
Both Feldman-Cousins and Bayes upper limits are smaller than the lower limit (1) on  $\epsilon_{eV}$ 
which was obtained  from the last NA64 bound \cite{NA64last} on visible dark photon decays.  
So we conclude that the use of the last NA64 bound (1) plus 
the experimental value (5) allows to 
exclude the $X(16.7)$ model with pure vector interaction at 90 \% C.L.

In general case  as a consequence of   the experimental value (5) and the formulae (3) we find  
\begin{equation} 
\epsilon^2_{Ve} - 5\epsilon^2_{Ae} = (-1.20 \pm 0.50 )\times10^{-6} \,.
\end{equation}
Consider at first the  case  of pure axial-vector interaction, i.e. $\epsilon_{Ve} =0$.
As a consequence  of (12) we find 
\begin{equation}
\epsilon_{Ae} = (0.49 \pm 0.09) \times10^{-3} \,.
\end{equation}
This value of $ \epsilon_{Ae}$   is excluded at 90 \%  C.L. level by the last NA64 bound (1)\footnote{ The NA64 bound (1) for nonzero $\epsilon_{Ae}$ and $m_{A^`} \gg m_e$  reads   
$\epsilon_e \equiv \sqrt{\epsilon^2_{Ve} + \epsilon^2_{Ae}} \geq 6.8 \cdot 10^{-4} $.} 
  . 
Consider now the case $|\epsilon_{Ve}| = |\epsilon_{Ae}| $  corresponding to the interaction of $X(16.7) $ 
with right-handed or left-handed electrons. 
For this case the  estimate for $\epsilon_{Ae}$ reads
\begin{equation}
\epsilon_{Ae} = (0.55 \pm 0.10) \times10^{-3} \,.
\end{equation}
while the NA64 bound for $\epsilon_{Ae}$ 
is
\begin{equation}
|\epsilon_{Ae}| = |\epsilon_{Ve}| \geq 0.48 \times10^{-3} \,.
\end{equation}
So we see that the model where  X(16.7) interacts  with left-handed or right-handed electron 
survives at 90 \% C.L. for $9.6 \cdot 10^{-4} \geq ~\epsilon_e \geq  ~6.8 \cdot 10^{-4}$ 
and moreover it explains  the $2.4\sigma$ discrepancy between the theory and the experiment for electron $(g_e-2)$ anomaly. 
The  NA64 experiment has to increase its sensitivity to $\epsilon^2$ parameter 
by factor 2  to discover or  reject the X(16.7) model with $V \pm A$ couplings.\footnote{For rather artificial 
case $g^2_{Ve} \approx 5g^2_{Ae}$ the contribution of the $X(16.7)$ boson to the $\Delta a_e$ is suppressed and the values 
$\epsilon_{Ve} \sim O(10^{-3})$ are not excluded.} 

Note that in ref.\cite{GK} a model with additional light vector boson  interacting with right-handed fermions 
of the first and second generations has been proposed. The model does not have $\gamma_5$-anomalies and the extension of the SM 
is possible for nonrenormalizable Yukawa interaction. For instance, for the electron and muon doublets 
 $L_e =(\nu_{eL}, e_L) $,   $L_{\mu} =(\nu_{\mu L}, \mu_L) $
the nonrenormalizable $SU_c(3) \otimes SU_L(2)\otimes U(1)\otimes U_R(1)$ gauge invariant 
Yukawa interaction reads
\begin{equation}
L_{Yuk}  = - \bar{h}_e\bar{L}_e\Phi H e_R  - \bar{h}_{\mu}\bar{L}_{\mu}\Phi H \mu_R    +   H.c.
\end{equation}
Here $\Phi$ is complex scalar field with   $<\Phi> \neq 0$ responsible for  
$X(16.7)$ vector boson and lepton  masses and $H= (H^+, H^0))$ is the Higgs isodoublet.  
For the $X(16.7)$ model in the unitaire gauge $\Phi = \Phi^*$, 
$H = (0, \frac{h}{\sqrt{2}} + <H>)$ neutral scalar field $\phi = 
\sqrt{2}(\Phi - <\Phi>)$ interacts with  electron  and  the effective Yukawa Lagrangian is
\begin{equation}
L_{eff,Yuk} = -h_e\bar{e}e \phi - h_{\mu}\bar{\mu}\mu \phi    \,,
\end{equation}
where $h_e = \frac{\bar{h}_e<H>}{\sqrt{2}}$,  $h_{\mu} = \frac{\bar{h}_{\mu}<H>}{\sqrt{2}}$.
The contribution of the scalar  field $\phi$  to the anomalous magnetic 
moment of electron is positive 
but it is negligible in comparison with the contribution (3). For anomalous magnetic moment of there are two contributions: 
the first one is negative contribution of the $X$ boson and the positive contribution of the $\phi$-scalar contribution.
The  Yukawa interaction (16) of the scalar field with muon
leads to   one loop contribution to  muon 
anomalous magnetic moment \cite{MOMENT}     
\begin{equation}
\Delta a_{\mu}(\phi) = \frac{h^2_{\mu}}{8\pi^2}\frac{m^2_{\mu}}{m^2_{\phi}}
\int^1_0\frac{x^2(2-x)dx}{(1-x)(1-\lambda_{\phi}^2x) + \lambda_{\phi}^2 x} \,,
\label{42}
\end{equation}
where $\lambda_{\phi} =  \frac{m_{\mu}}{m_{\phi}}$. 
For heavy scalar $m_{\phi} >> m_{\mu}$ 
\begin{equation}
\Delta a_{\mu}(\phi) = \frac{h^2_{\mu}}{4\pi^2}\frac{m^2_{\mu}}{m^2_{\phi}} [ln(\frac{m_{\phi}}{m_{\mu}}) 
- \frac{7}{12}] \,
\label{43}
\end{equation} 
and for light scalar $m_{\phi}  \ll m_{\mu} $  
\begin{equation}
\Delta a_{\mu}(\phi) = \frac{3h^2_{\mu}}{16\pi^2}  \,.
\label{44}
\end{equation}
One can find that that the scalar with a mass $m_{\phi} \approx 90~MeV$ cancels the negative 
$X$-boson contribution \cite{MOMENT}
\begin{equation}
\Delta a_{\mu}(X) = \frac{\epsilon^2_{Ve}\alpha}{2\pi}\frac{m^2_{\mu}}{m^2_{X}}
\int^1_0\frac{ 2x^2(1-x) +  2x(1-x)(x-4) - 4\lambda_{X}^2 x^3}{(1-x)(1-\lambda_X^2x) + \lambda_X^2 x} \,,
\label{45}
\end{equation}
and reproduces the correct value of the muon $g_{\mu} - 2$ 
anomaly \footnote{Here $\lambda_X = \frac{m_{\mu}}{M_X}$.} $\Delta a_{\mu} \approx 3 \times 10^{-9}$ \cite{MOMENT}.     

It should ve noted that for the scenario with dark photon decaying mainly into invisible particles 
last NA64 invisible bound \cite{NA64invisible} leads to the bound $\Delta a_e| \leq O(10^{-14})$ 
for dark photon, dark scalar particles with masses $m_X \leq  1~GeV$. For instance, for  $m_X = 16.7~MeV$  
NA64 bound on $\epsilon$ for invisible dark photon decays is  $|\epsilon| \leq 0.52 \cdot 10^{-4}$
that leads   to the bound on the anomalous electron magnetic moment   $|\Delta a_e|  \leq 1.9\cdot10^{-15}; 9.7\cdot10^{-15};   
 1.7 \cdot 10^{-14};   1.95 \cdot 10^{-14} $    for vector, axial vector, scalar and pseudoscalar mediators. It 
means that NA64 bounds for invisible mode on coupling constants of light hypothetical particles with electron are 
much stronger than the corresponding bound from anomalous magnetic moment $\Delta a_e$ for electron. 

Let us formulate our results. Last NA64 bound on $\epsilon$ parameter for visible dark photon decays 
plus the experimental value of the anomalous electron magnetic moment exclude $X(16.7)$ model with 
pure vector or axial vector interactions. The model with $V \pm A$ interaction still survives and moreover 
it can explain both the electron and muon $(g-2)$ anomalies.

\section*{Acknowledgments}
I am    indebted to S.N.Gninenko and V.A.Matveev  for useful discussions and comments.


\newpage





\begin{thebibliography}{99}


\bibitem{ATOMKI} A. Krasznahorskay  et al.(Atomki Collaboration), 
 Phys. Rev. Lett. {\bf V.116}(2016) 042501.
\bibitem{ATOMKI1} A. Krasznahorkay  et al.(ATOMKI Collaboration), 
arXiv:1910.10459v1 (2019).
\bibitem{Feng} J.L. Feng  et al., 
Phys.Rev. {\bf D95} (2017) 035017; \\
B. Fornal, Int.J.Mod.Phys. {\bf A32}  (2017)  1730020.
\bibitem{NA64} D. Banerjee et al.(NA64 Collaboration), 
Phys.Rev.Lett. {\bf 120} (2018) 231802. 
\bibitem{NA64last} D. Banerjee  et al.(NA64 Collaboration) CERN-EP-2019-284; arXiv:1912.1839. 
\bibitem{alpha} R.H.Parker et al., Science {\bf 360} (2018) 191.
\bibitem{13} T.Aoyama, T.Kinoshit, M.Nio, Phys.Rev. {\bf F97} (2019) 036001. 
\bibitem{15} D.Haneke, S.F.Hoogerheide, G.Gabrielse, Phys.Rev. {\bf A83} (2011) 052122.
\bibitem{Marciano} H.Davoudiasl,
 W.J.Marciano,  Phys.Rev. {\bf D98} (2018) 075011.
\bibitem{MOMENT} As a  review, see for example:
Dorokhov A.E., Radzhabov A.E., Zhevlakov A.S., 
 EPJ Web Conf. 2016. {\bf V.125} (2016) 02007; \\ 
Jegelehner F., Nyffeler A.   Phys. Rep. {\bf 477C} (2009) 1. 
\bibitem{Cousins} G.Feldman, R.A.Cousins, Phys.Rev. {\bf D57} (1998) 3873.
\bibitem{GK} S.N.Gninenko, N.V.Krasnikov, EPJ Web.Conf. {\bf 125} (2016) 02001; \\ 
arXiv1605.03056.
\bibitem{NA64invisible}Banerjee D. et al.(NA64 Collaboration),  Phys.Rev.Lett. {\bf 123} (2019) 
121801. 


\end{thebibliography}
\end{document}